\documentclass[11pt]{article}
\usepackage{blois2002,epsfig}
\def\be{\begin{equation}}
\def\ee{\end{equation}}
\def\bea{\begin{eqnarray}}
\def\eea{\end{eqnarray}}

\begin{document}
\vspace*{4cm}
\title{CP VIOLATION FROM DIMENSIONAL REDUCTION:\\ A SIMPLE EXAMPLE.}

\author{N. COSME}

\address{Service de Physique Th\'eorique -
 Universit\'e Libre de Bruxelles, \\ CP225 Bld du Triomphe,  1050 Brussels, Belgium.}

\maketitle\abstracts{
$CP$ is a symmetry of pure gauge theories, that is without scalar interactions. Actually its violation
originates in the Standard Model from the completely arbitrary Yukawa couplings. Thus, as the unification 
principle would wash out the arbitrariness of these couplings possibly relating them to the gauge interactions, 
the resulting theory should be expected as $CP$-invariant. According to this, a breaking mechanism of $CP$ should then
take place. We will explore here the possibility of $CP$ breaking through the dimensional 
reduction process.\cite{Cosme:2002zv}$ \; $ \cite{thirring} $ \; $\cite{autres}
}


Firstly, we resume briefly the issue of discrete symmetries in the context of dimensionaly extended space
or more particulary of parity in an even number of spatial dimensions.\cite{Gavela} In 3 spatial dimensions, the central
inversion $\overrightarrow{x} \rightarrow -\overrightarrow{x}$ and the specular reflexion, say 
$x_1 \rightarrow - x_1$, are two equivalent definitions of parity modulo a spatial rotation. However, for an
even number of spatial dimensions, the central inversion becomes simply an element of the rotation group, with $det=+1$.
The specular reflexion remains a discrete symmetry and the equivalence does not hold anymore. Actually, the best
generalisation of parity is the specular reflexion which leads to a generalisation of the $CPT$ theorem. This 
leads to a universal definition of parity whatever the number of dimensions: let take the reflexion of the last spatial
coordinate ($x_{d-1} \rightarrow - x_{d-1}$ for $d$ dimensions).

As a consequence, the scalar term $\bar{\psi}\psi$ in 4+1D is $P$-violating. Indeed, in 3+1D, it is easy to go
from $\bar{\psi}\psi$ to $\bar{\psi}i \gamma_5 \psi$ by a transformation on the semi-spinor and only the simultaneous
presence of both couplings achieve $P$-violation. In 4+1D, since the last component of a vector has to be identified with 
$\bar{\psi}i \gamma_5 \psi$ (with $i\gamma_5=\gamma_4$ to get a right extention of the Clifford algebra), the 
transformation cannot be done  and this results in $\bar{\psi}\psi$ to be $P$-violating while $CP$-conserving.
This is at the origin of $CP$ violation in the dimensional reduction process.

\section*{}

To the aim of getting those two components in a reduced theory, we consider a gauge theory in  4+1D:
\be \mathcal{L}= i\bar{\psi}D \!\!\!\!/\psi -M \, \bar{\psi} \psi, \ee
with $D_B= \partial_B -i e A_B$ ($B= 0,1,2,3,4$). We observe directly that the reduction to 3+1D could introduce 
effective complex mass term through non-vanishing contribution arising from the last component of the kinetic term, which
we denote by $X_4$: $\bar{\psi}(M+i \gamma_5 X_4) \psi$.

Such a structure could lead to $CP$ violation even if in the minimally coupled $U(1)$ case the complex phase can be removed 
in 3+1D. $CP$ violation can be achieved, first as it was considered by Thirring,\cite{thirring} using the $\partial_4$
term, identifying $X_4$ to the Kaluza-Klein mass $\frac{n}{R}$ and getting a non minimally coupled $U(1)$ from the 
reduction of gravity; or, as it is addressed here, taking an "expectation value" for the last component of the gauge field
$\langle A_4 \rangle\neq 0$ and extending the gauge group. This choice enables us to clearly separate $CP$ violation from
the use of exited KK states.

Let first comment on the possibility of a non-vanishing expectation value for $A_4$. As such, $\langle A_4 \rangle$ is not 
gauge invariant since the value of $A_4$ can be rotated away at any given point. A gauge invariant formulation is provided 
by a loop integral over a path which conserve 3+1D Lorentz invariance, i.e.:
\be X_4= \int dy \; A_4,\ee
assuming $X_4$ to be time and $\overrightarrow{x}$ independent ($y=x^4$). This quantity thus depends on the compactification scheme.

We turn now to a concrete example based on the $SU(2)$ group minimally coupled through the $W^a_A$ gauge bosons to fermions
with a mass, and assuming:
 \be \langle W_4 \rangle =\int dy \; W_4(y) =\left(\begin{array}{cc} w&  \\  &-w \end{array}\right).\ee
This results in the effective 3+1D Lagrangian:
\be\left(\begin{array}{cc}\bar{\psi_1} & \bar{\psi_2}\end{array}\right)i(\partial^\mu -i W^\mu_a \tau^a) \gamma_\mu
 \left(\begin{array}{c}\psi_1 \\ \psi_2\end{array}\right) 
+ \left(\begin{array}{cc}\bar{\psi_1} & \bar{\psi_2}\end{array}\right)\mathcal{M}
 \left(\begin{array}{c}\psi_1 \\ \psi_2\end{array}\right),\ee
where 
\be \mathcal{M}= \left(\begin{array}{cc} M+iw\gamma_5&  \\  &M-iw\gamma_5\end{array}\right).\ee
The mass matrix can be made real generally by a bi-unitary transformation, $\mathcal{M}'= U^\dagger_R \mathcal{M} U_L$.
In the present case, we can choose ($\alpha=\gamma_5 \arctan{\frac{w}{M}}$):
\be U_R=I, \qquad U_L=\left(\begin{array}{cc} e^{-i\alpha}&  \\  &e^{i\alpha} \end{array}\right).\ee

\begin{figure}\begin{center}
\epsfig{figure=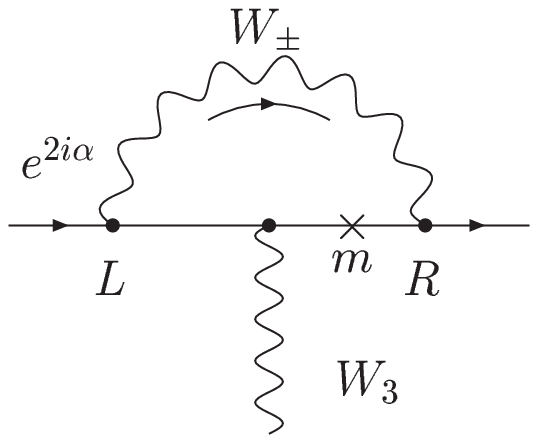,height=1.2in}
\caption{\label{fig:fig1}}
\end{center}
\end{figure}

The loop integral results in gauge symmetry breaking with two massive $W^\pm$ and one massless $W^3$. The use of such a
line integral has been developed in details by Hosotani in the framework of dynamical symmetry breaking.\cite{Hosotani} 
Even if the fermion masses are transformed to be real, the coupling to $W^\pm$ is no longer vectorial, but includes a 
phase between the left and right-handed part resulting in a $W^3$-dipole moment at one loop level (see Figure\ref{fig:fig1}).

$CP$ violation thus arises here via the fermion mass matrix as in the Standard Model and its origin is related to both 
dimensional reduction and breaking of the internal symmetry. However, strong limitations need extention to more realistic
models, namely: mass degenerancy, the vectorlike coupling in the current basis, etc.

\section*{}

For instance, since in 4+1D only vectorlike couplings are allowed, to get chiral couplings we could consider topological 
defect, e.g. domain wall,\cite{Rubakov} which selects chiral massless fermionic modes in its core. Moreover, since $CP$ violation arises
here through the generation of fermion mass from the gauge bosons, we choose the defect to be part of the internal group resulting
in the selection of left and right-handed fermions which are linked through the loop integral. Nevertheless, this procedure
divides directly the number of fermionic components by two and thus implies for the $U(1)$ case no complex mass while
for the  $SU(2)$ case one single complex mass which can be safely redefined as real.

One simple example is then provided by $SU(4)$. Let us consider the vaccum configuration:
\be \Phi = \frac{\phi(y)}{2} \left(\begin{array}{cccc}
1 &  &  & \\ 
 & 1 &  & \\ 
 &  & -1 & \\
& & &-1
\end{array}\right),\;
\chi = \frac{\langle \chi \rangle}{2} \left(\begin{array}{cccc}
0 &  &  & \\ 
 & 0 &  & \\ 
 &  & 1 & \\
 &  &  &-1
\end{array}\right),\;
 \eta = \frac{\langle \eta \rangle}{2} \left(\begin{array}{cccc}
1 &  &  & \\ 
 & -1 &  & \\ 
 &  & 0 & \\
 &  &  &0
\end{array}\right),\ee
with $\Phi$ being the domain wall from which the fermion localisation will select an $SU(2)_L \times SU(2)_R \times U(1)_A$ group
between fermion zero modes: $\left(\begin{array}{cccc} u^1_L & d^1_L  & u^2_R  & d^2_R \end{array}\right)$; while $\eta$ and $\chi$,
 acquiring a constant vev, will break down respectively the $SU(2)_L$ and the $SU(2)_R$ subgroup. 
After that, the generation of fermion masses needs non-diagonal scalars $H^1$ 
and $H^2$; e.g.:
\be H^1 = \frac{\langle H^1 \rangle}{2} \left(\begin{array}{cccc}
 0& 0 & 1 & 0\\ 
 0& 0 & 0 &0 \\ 
 1& 0 & 0 &0 \\
0&0 &0 &0
\end{array}\right)=\langle H^1 \rangle \lambda_4 ,
 \qquad H^2 = \frac{\langle H^2 \rangle}{2} \left(\begin{array}{cccc}
0 & 0 & 0 &0 \\ 
0 & 0 & 0 &1 \\ 
0 & 0 & 0 & 0\\
 0&  1&  0&0
\end{array}\right)=\langle H^2 \rangle \lambda_{11}.\ee

Since those two breakings commute, they clearly minimise their interaction potential, but moreover allow the 
Hosotani term to get a component in each direction without cost of energy:
$\int dy\;  W^4 = w_4 \lambda_4 + w_{11} \lambda_{11}.$
This feature provides actually two masses with two phases, i.e.:
\be \bar{u}^1_L \;\;\frac{1}{2}(m_1 \langle H^1 \rangle +i w_4 \gamma_5) \;\;u^2_R + 
\bar{d}^1_L \;\;\frac{1}{2}(m_2 \langle H^2 \rangle +i w_{11} \gamma_5)\;\; d^2_R + h.c.. \ee

It can be easily checked that both phases cannot be removed completely from the Lagrangian. 
However, to have $CP$ violation through the $W_L$ alone, more generations are needed.
The model discussed here is not yet realistic in that charge assignations in the fundamental of 
$SU(4)$ are not compatible with the observed ones.

\section*{Acknowledgments}
 This work was done in collaboration with J.-M. Fr\`ere and L. Lopez Honorez and is supported in part by IISN,
la Communaut\'{e} Fran\c{c}aise de Belgique (ARC), and the belgian federal governement (IUAP).

\end{document}